# Evolutionary accumulation modelling in AMR: machine learning to infer and predict evolutionary dynamics of multi-drug resistance

Jessica Renz[1], Kazeem A. Dauda[1], Olav N. L. Aga[2,3], Ramon Diaz-Uriarte[4,5], Iren H. Löhr[3,6], Bjørn Blomberg[3,7,8], Iain G. Johnston[1,2,*]

1. Department of Mathematics, University of Bergen, Bergen, Norway
2. Computational Biology Unit, University of Bergen, Bergen, Norway
3. Department of Clinical Science, University of Bergen, Bergen, Norway
4. Department of Biochemistry, School of Medicine, Universidad Autónoma de Madrid, Madrid, Spain
5. Instituto de Investigaciones Biomédicas Sols-Morreale (IIBM), CSIC-UAM, Madrid, Spain
6. Department of Medical Microbiology, Stavanger University Hospital, Stavanger, Norway
7. Department of Medicine, Haukeland University Hospital, Bergen, Norway
8. National Advisory Unit for Tropical Infectious Diseases, Haukeland University Hospital, Bergen, Norway
* correspondence to iain.johnston@uib.no

## Abstract

Can we understand and predict the evolutionary pathways by which bacteria acquire multi-drug resistance (MDR)? These questions have substantial potential impact in basic biology and in applied approaches to address the global health challenge of antimicrobial resistance (AMR). Here, we review how a class of machine learning approaches called evolutionary accumulation modelling (EvAM) may help reveal these dynamics using genetic and/or phenotypic AMR datasets, without requiring longitudinal sampling. These approaches are well-established in cancer progression and evolutionary biology, but currently less used in AMR research. We discuss how EvAM can learn the evolutionary pathways by which drug resistances and AMR features are acquired as pathogens evolve, predict next evolutionary steps, identify influences between AMR features, and explore differences in MDR evolution between regions, demographics, and more. We demonstrate a case study on MDR evolution in *Mycobacterium tuberculosis* and discuss the strengths and weaknesses of these approaches, providing links to some approaches for implementation.

## Introduction

Anti-microbial resistance (AMR) occurs when pathogens (like bacteria) evolve resistance to the drugs we use to kill them. Multi-drug resistance (MDR) occurs when a pathogen evolves resistance to multiple groups of drugs (usually three or more clinically relevant drug types). AMR is a growing health issue across the globe, particularly in developing countries: in 2019 and 2021, 1.27m and 1.14m deaths respectively were attributable to bacterial AMR, with the highest rates in sub-Saharan Africa (Murray et al., 2022; Naghavi et al., 2024). AMR results in large excess societal costs due to prolonged illness and hospitalizations, need for isolation and use of expensive reserve medicines, and patients' loss of productivity (Dadgostar, 2019; Morel et al., 2020; R. R. Roberts et al., 2009). AMR is one of the World Health Organization's top 10 threats to human health (World Health Organisation, 2021).

Given the importance of AMR for global health (Aslam et al., 2021; Djordjevic et al., 2024), many research programmes are investigating different aspects of the phenomenon (T. R. Walsh et al., 2023). These include basic science questions: from molecular mechanisms by which pathogens resist drug treatments (Darby et al., 2023; Ray et al., 2017; Reygaert, 2018; Sullivan et al., 2020) and surveillance of pathogens (Baker et al., 2023; Djordjevic et al., 2024), through the dynamics by which these mechanisms evolve in pathogen populations (Baquero et al., 2021; Maeda et al., 2020; Nichol et al., 2015). Clinically-motivated questions include the development of new treatments (Piddock, 2012; Terreni et al., 2021; C. Walsh, 2003), strategies for prevention ranging from infection prevention to antibiotic stewardship (Septimus, 2018; Spellberg et al., 2016), improved diagnostics, vaccination (Laxminarayan et al., 2024), as well as population education and citizen science (A. P. Roberts, 2020). Broader areas of investigation include the interplay with animal populations and the environment in a One Health Perspective (Hetland et al., 2024; Morel et al., 2020), and the public perception of AMR as a threat (Duncan et al., 2020; Fimreite et al., 2024). This range of research activity has emphasized the power of interdisciplinary approaches in combatting AMR, and of quantitative approaches to extract insights from large-scale data (Djordjevic et al., 2024; Holt et al., 2015; Munk et al., 2022).

Here, we focus on one particular aspect of AMR – the dynamics by which multidrug resistance evolves in pathogens – and a particular emerging interdisciplinary approach to understand and predict these dynamics. This approach is called "evolutionary accumulation modelling" or EvAM (Diaz-Uriarte & Herrera-Nieto, 2022; Diaz-Uriarte & Johnston, 2024; Schill et al., 2024).

The core task of EvAM is to use data to learn the pathways by which an evolving system accumulates features over time. These features (usually called "characters" in evolutionary biology) are typically binary representations of particular properties or values of interest in the system. We have, for now, deliberately left these descriptions broad, because EvAM has been applied across a very broad range of contexts in biology and beyond (Diaz-Uriarte & Johnston, 2024). Examples include the accumulation of mutations in developing tumours (Angaroni et al., 2022; Beerenwinkel et al., 2015; Gerstung et al., 2009; Hjelm et al., 2006; Luo et al., 2023; Montazeri et al., 2016; Nicol et al., 2021; Ross & Markowetz, 2016; Schill et al., 2020); the accumulation of gene losses in evolving organelles across species (Johnston & Williams, 2016); the accumulation of symptoms in disease progression across patients (Johnston et al., 2019); the accumulation of completed tasks in online courses across students (Peach et al., 2021); and indeed the evolution of drug resistance in pathogens (Aga et al., 2024; Beerenwinkel, Däumer, et al., 2005; Greenbury et al., 2020; Moen & Johnston, 2023; Montazeri et al., 2015; Posada-Céspedes et al., 2021). The dynamics of any evolving system where a given observation can be represented by a collection of presence/absence markers can be analysed using EvAM. The mathematical structure of the "space" of such binary marker strings, and evolutionary steps between them, is a hypercube – the analogy of the familiar 3D cube in any number of dimensions – leading to several approaches being classed under "hypercubic inference".

Given observations of such a system, different EvAM approaches perform machine learning to estimate different interesting properties of the underlying accumulation process. For example, does acquiring feature A make it more or less likely that feature B will be acquired? If a given instance has features A, B, and C, what is the next most likely feature to be acquired? Are there combinations of features that correlate with future behaviour – like death or survival of a patient with those symptoms? And are these answers universal, or do they differ across patients, hospitals, regions, or countries? Importantly, longitudinal data is not required by EvAM to learn dynamic behaviour – information on the orderings of, and influences between, features can be inferred from cross-sectional observations.

## Evolution accumulation modelling for MDR evolution

How can EvAM be useful in the study of MDR? First consider, for example, the case where our features are resistances to each of a set of drugs, and our observations are a collection of pathogen isolates. Then each of the questions above becomes potentially valuable both for basic biology and for clinical strategy. If resistance to drug A makes an isolate more likely to evolve resistance to drug B, maybe we should avoid that combination. But if drug A decreases the probability of drug B resistance evolving, the combination therapy may be more effective. If a new isolate is observed in the clinic that is resistant to drugs A, B, and C, then predicting the next resistance "step" suggests a treatment that should be avoided. If particular combinations of drug resistances correlate with, for example, virulence, a triaging strategy can be designed. And if differences in the evolutionary dynamics of MDR (for example, between countries, or patient demographics) can be identified and characterized with data, interventions that are more targeted to a particular circumstance may be designed and tested.

As a more concrete example, several EvAM approaches have been used by our group to study the evolution of MDR in a Russian *Mycobacterium tuberculosis* dataset (Casali et al., 2014) that has become a benchmark for the approaches that we develop. Outlined in the next section, this example case study involves features describing resistance or susceptibility to each of 10 drugs across a dataset involving 1000 sequenced isolates. The different ordering of drug resistance acquisitions, influences between them, predicted future behaviours, and similarities and differences between countries, have been learned using different EvAM approaches (Fig. 1; (Aga et al., 2024; García Pascual et al., 2024; Greenbury et al., 2020; Moen & Johnston, 2023)).

The features under study in EvAM need not only be phenotypic. Genetic features can readily be studied too, addressing research questions accordingly. AMR typically evolves through mutations and horizontal gene transfer events. While a particular AMR phenotype can be the result of a number of different genetic events, mapping the actual genetic feature strengthens the model by providing a more direct evidence of the ongoing evolution. For example, SNPs or other small-scale genetic changes linked to resistance can be considered, given targeted or whole-genome sequencing data. Does the acquisition of one AMR-linked mutation influence the acquisition rate of others?

On another scale – and features at different scales can readily be combined in EvAM – the presence or absence of whole genes conferring resistance can be considered. Does the presence of a particular gene, for example, influence the propensity of a strain to acquire other AMR-linked changes? Indeed, other instances of EvAM used to model AMR include estimation of evolutionary pathways to drug resistance in HIV (Beerenwinkel, Däumer, et al., 2005; Posada-Céspedes et al., 2021). Other AMR work with an implicit or explicit EvAM picture has included "bottom-up" pictures coupling an EvAM-like hypercubic evolutionary space with fitness landscape values in *Escherichia coli* (Das et al., 2020; Das & Krug, 2022; Nichol et al., 2015; Tan et al., 2011), that also address beta-lactamase mutants (Farr et al., 2023). These empirically-linked landscapes often consider a limited number of features (by necessity, as the space of genotypes expands exponentially with the number of features considered). By contrast, the EVaM approaches we will consider take a "top-down" perspective, less directly coupled to fitness and more concerned with the occurrence of, and transitions between, different states across observation sets.

## An EvAM MDR example: multidrug resistance in tuberculosis

To illustrate an EvAM application to MDR, consider the tuberculosis dataset mentioned above. Here, resistance profiles describe the presence or absence of resistance to ten different drugs (Fig. 1A). 395 of 1000 isolates have complete profiles; others are missing at least one drug entry. A phylogeny, constructed in the original study using independent genetic information from the isolates, is also available (Fig. 1A).

To use EvAM here, each isolate with a complete profile is assigned a length-10 binary string, with a 0 in a given position meaning "susceptible to this drug" and a 1 meaning "resistant to this drug". In this illustrative example, we assume that resistances are acquired irreversibly. We then reconstruct ancestral states with a maximum parsimony picture. For irreversible dynamics, this means an ancestor is assumed to have been resistant (1) if both descendants are resistant, or susceptible (0) if neither or only one descendant is resistant. The collection of ancestor-descendant transitions throughout the phylogeny is used as the input data for EvAM.

We can then use EvAM approaches to infer the evolutionary pathways of MDR across these observations. First, we consider the "transition network" learned from the data using HyperTraPS-CT (Aga et al., 2024). This network describes the likely evolutionary steps that an isolate with a particular drug resistance will next take (Fig. 1B). The network structure shows clear canalization – a limited number of pathways are supported, with some variability (branching in the network). For example, isoniazid (INH) and streptomycin (STR) resistances are typically the first to be acquired, often followed by rifampicin (RIF), then ethambutol (ETH), prothionamide (PRO), and pyrazinamide (PZA), although the particular orderings of resistances are flexible with this general theme. WHO guidelines suggest INH and RIF treatment, supplemented with PZA and with ETH in regions of high INH resistance (World Health Organization, 2022) – so these orderings may in part reflect a response to treatment regimens, while in part suggesting dynamics (like PRO resistance) that cannot immediately be predicted from such protocols (see Discussion).

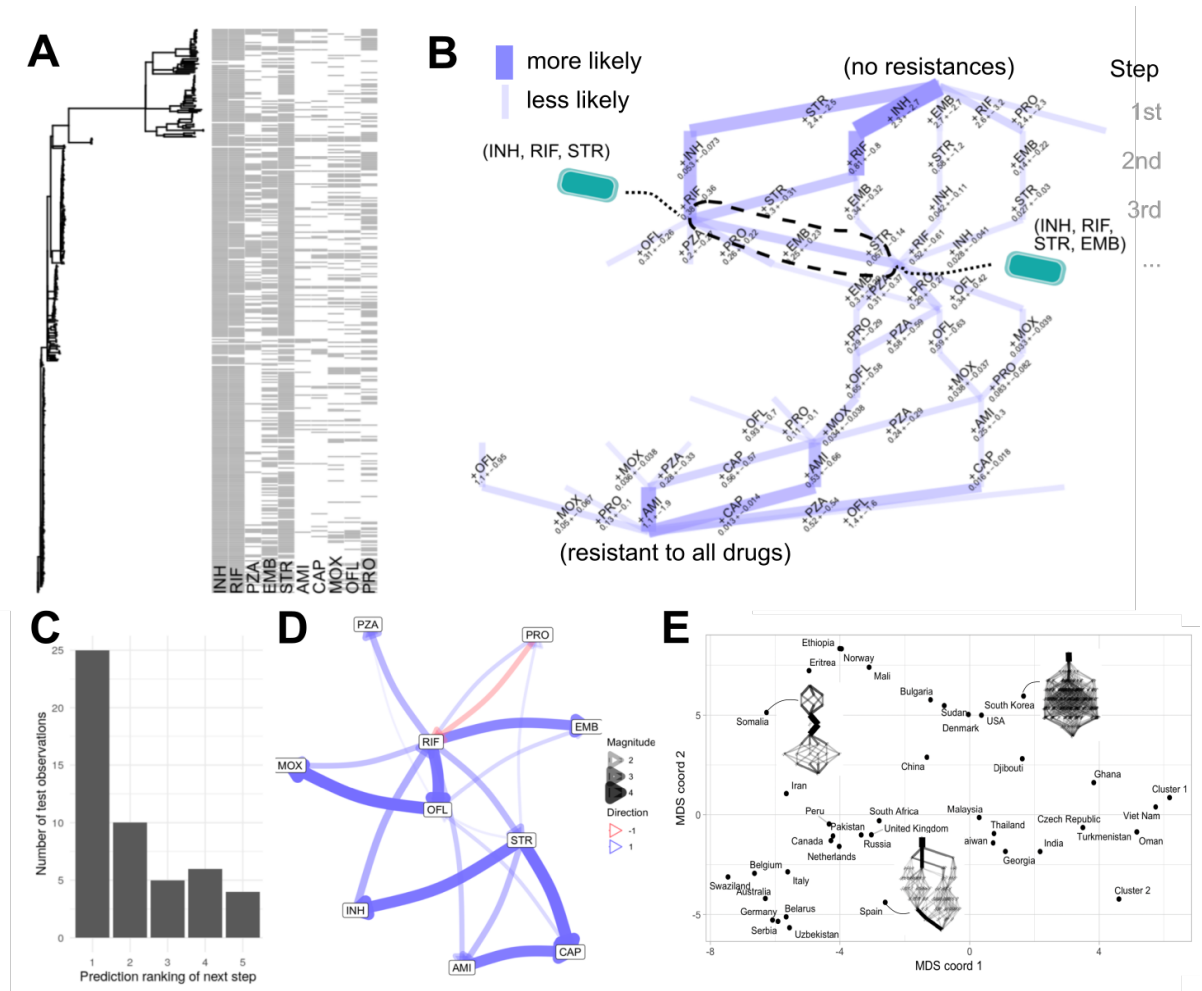

Figure 1. **Example insights from EvAM for MDR evolution.** These results are from HyperTraPS-CT and HyperHMM applied to MDR evolution in tuberculosis (Aga et al., 2024; Casali et al., 2014; García Pascual et al., 2024; Moen & Johnston, 2023; Olson et al., 2023)). (A) Dataset, comprising a set of phylogenetically-linked isolates with drug resistance profiles (grey, resistant; white, susceptible) to each of ten drugs (codes in text). (B) Inferred hypercubic transition network describing likely pathways of MDR evolution from an initial, fully susceptible state (top) to a fully resistant state (bottom). Each step down the network corresponds to the acquisition of resistance to another drug; line width gives probability of that step. Central graphic with dotted lines illustrates a prediction: from a state with INH, RIF, STR resistance, the predicted most likely next step is EMB resistance. (C) Testing predictions: when a prediction is made about the next likely step (as in (B)), is that step actually the next that occurs in reality? Horizontal axis gives the predicted ordering of the step that was, in reality, the next step. (D) Interactions between features. Inferred positive and negative influences: does resistance to drug A increase (blue), or decrease (red), the probability of acquiring resistance to drug B? (E) Comparison of inferred dynamics across countries. Multidimensional scaling plot of inferred transition networks for different countries (clusters 1 and 2 are sets of countries with identical observations), illustrating the structure of between-country similarities and differences in dynamics (García Pascual et al., 2024). Insets give some example networks, styled as in (B). Drug codes: INH (isoniazid); RIF (rifampicin, rifampin in the United States); PZA (pyrazinamide); EMB (ethambutol); STR (streptomycin); AMI (amikacin); CAP (capreomycin); MOX (moxifloxacin); OFL (ofloxacin); and PRO (prothionamide).

The transition network provides direct predictions about the future behaviour of a given strain. For example, given an isolate currently resistant to isoniazid, rifampicin, and streptomycin, what drug resistance will most likely be acquired next? As Fig. 1B shows, the trained model suggests that ethambutol resistance is the most likely next step (probability 29.2%), although others are possible. The performance of these predictions can be tested with a training-test split. Here, a subset of the data is used to train the model, and an unused subset is used to test predictions. By far the most common outcome is that the true next step in the test data was the one predicted by the trained model (Fig. 1C).

Next, we can explore interactions between features – whether one drug resistance makes it more or less likely to evolve another. Fig. 1D shows the influences inferred from the data. Several positive influences are detected – for example, RIF resistance increases the probability of acquiring ofloxacin (OFL) resistance, which in turn strongly increases the probability of moxifloxacin resistance (MOX). OFL and MOX are both fluoroquinolones, suggesting a biochemical rationale for this latter influence. Perhaps more interestingly, a negative interaction is also detected: PRO resistance makes RIF resistance less likely, suggesting rifampicin as a choice for strains that are prothionamide-resistant.

We can also use EvAM to learn evolutionary dynamics using data sourced from different cases – for example, from different countries. In Fig. 1E we used HyperHMM (Moen & Johnston, 2023) to infer transition networks using data on tuberculosis resistance from the Bacterial and Viral Bioinformatics Resource Centre (BV-BRC) (Davis et al., 2016; Olson et al., 2023) for different countries around the world (García Pascual et al., 2024). Here (methods taken from (García Pascual et al., 2024)) we retained records for the nine drugs (ofloxacin, ethambutol, isoniazid, streptomycin, capreomycin, rifampin, kanamycin, amikacin, pyrazinamide) for which > 100k observed isolates were available, subsetted by country of origin. We discarded records with any missing data for these drugs, leaving a total of 117207 isolates. In this illustrative case, we retain only unique drug resistance patterns within each country, to guard against pseudoreplication (see Discussion), retaining between 1 and 93 unique patterns for each of 53 countries. Following inference with HyperHMM, the different transition networks are compared using the "$l_1$ norm" – a summed absolute difference between transition probabilities. This is visualised using multidimensional scaling in Fig. 1E, so that more similar transition networks generally appear closer than more different ones. The different sources of variability – different first-step drug resistances, different subsequent canalized pathways, and different amounts of variability in the dynamics – can then be interrogated across country sets, as described further in (García Pascual et al., 2024).

This case study demonstrates several outputs of EvAM approaches, and also some questions which can arise. For example, are these inferred dynamics an intrinsic property of the bacterium itself, a response to the specific dosing regimen used in Russia, or a combination? To what extent can this trained model be used to predict pathogen evolution in other pathogens, other countries, or other settings? How could horizontal transfer of resistance traits (for instance, transfer of genes via plasmids) be accommodated? In the discussion we address these and other points in the interpretation of EVaM approaches.

## What is EvAM not?

Substantial exciting quantitative research is currently using large-scale data to explore AMR evolution and behaviour (Holt et al., 2015; Munk et al., 2022; Wyres et al., 2020). Having attempted to describe what EvAM is, and some ways that it may be useful in MDR evolution contexts, it will be important for us to describe what EvAM is *not*.

EvAM is not (directly) a **clustering or dimensionality reduction method**. A common early step in large-scale data analysis – including genomic AMR data -- is to use clustering and/or projection methods to reduce dimensionality (including principal components analysis (PCA); see, for example, (Munk et al., 2022)). EvAM does not directly do this, although it would be possible to using clustering methods to construct effective features for EvAM analysis (see Discussion). Rather, EvAM considers a given set of features, and how they evolve over time.

EvAM is not (directly) a **classification tool** for assigning categories to genomic (or other data). In AMR, many methods exist for – for example – inferring phenotypic drug resistances or other profiles from genotypes: some examples are (Davis et al., 2016; Hyun et al., 2020; Kavvas et al., 2020) and several are compared in (Ren et al., 2022). EvAM does not do this, instead working with features that it is provided as input (which can include such inferred profiles). However, EvAM can be used to assign likely evolutionary trajectories to given samples, which has been used in illustrative computation of triage high versus low-risk disease states (Johnston et al., 2019).

EvAM is not (exclusively) an **approach for estimating phylogenetic relationships** between isolates or observations. Some EvAM approaches either use the observations themselves, or independent information, to estimate a phylogeny as an intermediate step (Aga et al., 2024; Greenbury et al., 2020; Johnston & Williams, 2016; Luo et al., 2023). But the goal of EvAM is usually to learn how features evolve over a (given or estimated) phylogeny. In this sense, it resembles the so-called Mk model in phylogenetic comparative methods (Johnston & Diaz-Uriarte, 2024). The resemblance has some nuances, discussed in "Implementation" below. But EvAM typically assumes either independent, cross-sectional observations (likely not appropriate in AMR settings) or requires an estimated phylogeny and an approach for ancestral state reconstruction to be provided. If no phylogeny is available, progress can still be made by placing bounds on the possible dynamics (see Discussion).

EvAM is not a **microscopic quantitative model** of how changes arise and accumulate. Connecting the outputs of EvAM even to a model of fitness is non-trivial (Baquero et al., 2021; Diaz-Colunga & Diaz-Uriarte, 2021; Diaz-Uriarte, 2018; Diaz-Uriarte & Johnston, 2024; Diaz-Uriarte & Vasallo, 2019; Misra et al., 2014). Several exciting studies have successfully coupled a hypercubic EvAM space to direct fitness values, leading to experimental and theoretical insight (Nichol et al., 2015; Tan et al., 2011). However, more detailed insights about molecular mechanisms do not immediately arise from

EvAM output. Given case-specific information, the outputs of EvAM can readily be used in hypothesis testing about generative mechanisms (Johnston & Røyrvik, 2020; Johnston & Williams, 2016; Williams et al., 2013).

## Implementations and source data

Most EvAM methods developed in the last decade have free open-source software distributions available on repositories like Github. The meta-platform EvAM-Tools[1] (Diaz-Uriarte & Herrera-Nieto, 2022) brings a collection of these together with an easy-to-use interface. Many are implemented in R (R Core Team, 2022), requiring some familiarity with how to load libraries and data in an R environment. In most cases, a matrix or dataframe of binary-string observations can be directly passed to a function in the codebase.

However, the particular evolutionary setting of most MDR problems makes several established approaches less applicable. Many EvAM approaches traditionally assumed that observations are independent, cross-sectional instances of the process under study, which involves irreversible acquisition of features over time. This picture originates from the historical focus on cancer progression – where observations were typically tumour profiles from independent patient samples. But treating phylogenetically-linked observations as independent leads to the problem of pseudoreplication (Boyko & Beaulieu, 2023; Maddison & FitzJohn, 2015; Revell, 2010; Rohle, 2006; Schraiber et al., 2024; Uyeda et al., 2018). Here, because features may be inherited from an ancestor rather than independently acquired, an independent picture may give undue statistical weight to observations that actually reflect commonality by descent.

The true phylogeny connecting observations in an evolving tumour is usually unavailable. This makes the workflow in cancer studies slightly different from that in evolutionary biology (including the evolution of AMR) where an independently constructed phylogeny is often available. Several EvAM methods from the cancer literature use cross-sectional data to estimate the ordering of feature acquisitions that give rise to an observed set of observations, recently including Mutation Order[2] (Gao et al., 2022), SPhyR[3] (El-Kebir, 2018), SCITE[4] (Jahn et al., 2016), SiFit[5] (Zafar et al., 2017), with approaches generalising these pictures including REVOLVER[6] (Caravagna et al., 2018) and HINTRA[7] (Khakabimamaghani et al., 2019). Many methods above make the "infinite sites" assumption – effectively, that a given change can only ever occur once -- which then allows us to reconstruct both the phylogeny and the ancestor states in a so-called "perfect phylogeny". However, this picture neglects the possibility that the same feature may arise several times independently (parallel evolution), and so may not be best suited to an AMR perspective. SCARLET[8] (Satas et al., 2020) allows loss of features as well as acquisitions, substantially generalizing the supported dynamics. Other cancer approaches more generally attempt to infer dependencies between features from cross-sectional data, so that, for example, feature A is required for the acquisition of feature B. Parallel approaches generalize this to a stochastic picture, where feature A influences the (random) acquisition of feature B, including mutual hazard networks (MHN)[9] (Schill et al., 2020).

---

[1] https://github.com/rdiaz02/EvAM-Tools
[2] https://github.com/lkubatko/MO
[3] https://github.com/elkebir-group/SPhyR
[4] https://github.com/cbg-ethz/SCITE
[5] https://github.com/KChen-lab/SiFit
[6] https://github.com/caravagnalab/revolver
[7] https://github.com/sahandk/HINTRA

One set of EvAM approaches – under the umbrella of "hypercubic inference" – were designed to deal with phylogenetically-embedded data, including from MDR. These include HyperTraPS (Greenbury et al., 2020; Johnston & Williams, 2016), HyperHMM[10] (Moen & Johnston, 2023), and HyperTraPS-CT[11] (Aga et al., 2024). These approaches take transitions – from an ancestral to a descendent state -- as fundamental observations. Such transitions can be read off from a phylogeny when combined with a preliminary step of ancestral state reconstruction, for example by assuming rare irreversible transitions as with the tuberculosis case above. Other approaches including TreeMHN[12] (Luo et al., 2023) have been developed to analyse "phylogenetic" data from longitudinal studies within tumours, and may be coerced to handle phylogenetic data in an AMR context. Recently, the Mk model from phylogenetic comparative methods has been used in an EvAM setting, simultaneously performing ancestral state reconstruction and inference of feature dynamics (HyperMk[13], (Johnston & Diaz-Uriarte, 2024)). This approach also – rarely among EvAM methods -- supports reversible dynamics, potentially making it easier to capture MDR feature that are acquired reversibly (for example, through the acquisition of plasmids or other transferable genetic elements which can then be lost). However, it is computationally expensive and limited to a small number of features compared to other approaches.

To summarise: not all EvAM approaches from the cancer literature support the combination of phylogenetically-linked data, stochastic positive and negative dependencies, non-additive influences of sets of features, and reversibility that may be appropriate in AMR cases. TreeMHN supports the first two and HyperTraPS(-CT) and HyperHMM support the first three points (if ancestral state reconstruction can be performed). HyperMk is, to our knowledge, the only approach that supports all these points (including reversibility), but at the moment is limited in the number of features it supports.

To demonstrate the implementation of EvAM from source data, we have created a Github repository at https://github.com/StochasticBiology/EvAM-MDR with the code necessary to download, preprocess, and analyse the data from the tuberculosis case study above using HyperTraPS-CT. Once libraries, dataset, and tree are imported (Fig. 1A), training the model is accomplished in two lines of R code, and producing each plot for visualization (Figs. 1B, D) is another line of code. Complete code for the analyses in Fig. 1 can be found in the repositories for HyperTraPS-CT (Aga et al., 2024) and the "weight-filtration comparison curve" (WFCC) method for comparing EvAM outputs (García Pascual et al., 2024).

---

[8] https://github.com/raphael-group/scarlet
[9] https://github.com/RudiSchill/MHN
[10] https://github.com/StochasticBiology/hypercube-hmm
[11] https://github.com/StochasticBiology/hypertraps-ct
[12] https://github.com/cbg-ethz/TreeMHN
[13] https://github.com/StochasticBiology/hypermk

## Discussion and future directions

As with any approach, different EvAM methods have strengths and weaknesses; some have been outlined above. There is a substantial collection of research directions that have potential to expand and improve the use of these approaches in AMR evolution. Three topics that, in our experience, are the most common questions from an applied perspective involve reversibility (particularly pertinent when horizontal gene transfer is an important mechanism for AMR evolution), selective pressures, and continuous values. We will discuss these in turn.

As we have mentioned, many existing EvAM approaches assume that features are acquired irreversibly. In some cases this is not necessarily an unreasonable picture. Many of the AMR-conferring mutations in the tuberculosis case above, for example, are chromosomal changes, which may have a reasonably long persistence time in strains (Casali et al., 2014). However, in other cases, reversibility is likely an important aspect of the evolutionary dynamics. In *Klebsiella pneumoniae*, for example, many AMR genes are acquired via plasmids or other horizontally transferred genetic elements, which can readily be lost (especially if selective pressures due to drugs are removed) (Holt et al., 2015). Traditional EvAM approaches, dealing with irreversible accumulation, cannot capture such loss dynamics. One reason is that reversibility means that, in principle, an infinite set of evolutionary pathways can exist. For example, if we see a pathogen with resistance to drugs A and B, an irreversible picture would propose "gain A, gain B" and "gain B, gain A" as the two possible pathways. A reversible picture would allow any number of losses of either resistance following a gain, challenging the learning process. The recent work mentioned above, using the Mk model in an EvAM setting, can address this, but further work is needed to overcome its currently limited scale (Johnston & Diaz-Uriarte, 2024).

The next question, selective pressure, has several subquestions. First, as posed in the tuberculosis case study above – how can we unpack intrinsic, "universal" evolutionary behaviour from the response to the particular set of selective pressures found in a given dataset (for example, the particular drug use regimen in a given country)? Second, can the evolutionary dynamics inferred from a given dataset be interpreted as a fitness landscape or other quantitative picture directly linked to population behaviour?

The first subquestion may be addressed using observations taken from different cases where the selective pressures facing pathogens are different. For example, different countries may employ different drug regimens against a given pathogen. Comparing inferred evolutionary dynamics across such instances will reveal the similarities (corresponding more to intrinsic behaviour) and differences (corresponding more to case-specific responses), which can be quantified with summary statistics. If the inferred ordering of resistance acquisitions in each country always exactly follows the dosing regimen in that country, MDR dynamics are likely purely determined by dosing exposure. But if (as in the tuberculosis case above), there are observed departures from strict agreement with dosing regimen, other factors – possibly intrinsic to the pathogen's biology, or other environmental effects – likely also shape evolution. This is particularly likely if resistance machinery comes with a high cost in the absence of drugs. Many EvAM approaches explore between-case variability by inferring different models from different subsets of the data (Angaroni et al., 2022; Caravagna et al., 2018; Johnston et al., 2019; Johnston & Røyrvik, 2020; Nicol et al., 2021; Williams et al., 2013); we have developed some early approaches to compare EvAM outputs across different observation sets, including simple PCA (Williams et al., 2013), probabilistic summaries (Dauda et al., 2024), and the more detailed WFCC approach borrowing from topological data analysis (Fig. 1E; (García Pascual et al., 2024)). EvAM approaches learning deterministic tree-based dynamics have employed mixtures of trees to account for subject heterogeneity, including in drug-resistance studies of HIV (Beerenwinkel et al., 2004; Beerenwinkel, Rahnenführer, et al., 2005; Bogojeska, Alexa, et al., 2008; Bogojeska, Lengauer, et al., 2008), though to our knowledge these approaches have not been applied to phylogenetically-linked data or models involving stochastic dependencies. Ongoing research is considering gene-gene-environment (GxGxE) interactions in AMR and their influence on predictability (Das et al., 2020). In future, other approaches perhaps including inspiration from mixed Markov models (Fridman, 2003) may prove valuable in quantifying such between-case comparisons.

The second question – a connection to fitness landscapes – is an ongoing area of research in EvAM (Diaz-Uriarte & Johnston, 2024). In an AMR context, hypercubic pictures of drug resistance spaces and associated fitness landscapes have been used to guide experimental treatment strategies (Nichol et al., 2015) and explore evolutionary trajectories (Tan et al., 2011). It is not in general straightforward to link inferred evolutionary dynamics to a specific fitness landscape; noise, epistasis, and other issues complicate the task (Baquero et al., 2021; Diaz-Colunga & Diaz-Uriarte, 2021; Diaz-Uriarte, 2018; Diaz-Uriarte & Johnston, 2024; Diaz-Uriarte & Vasallo, 2019; Misra et al., 2014). AMR-specific theory for learning and exploiting information about fitness landscapes is a current area of research (Baquero et al., 2021), and linking this picture to EvAM outputs is an exciting avenue of future research to make these connections more general and quantitative.

The inclusion of continuous values in EvAM is pertinent for AMR applications because many quantities of interest are not simple presence/absence factors. Resistance phenotypes, for example, are often fundamentally reported as (continuous) minimum inhibitory concentrations (MICs) for a drug – with a threshold then applied to convert this to a susceptible/resistant factor. Previous EvAM work has had some success binarizing variables using a collection of inequalities (for example, is $x < 0$, $x < 1$, $x < 2$ and so on), effectively imposing different thresholds (Johnston et al., 2019). Recent developments in phylogenetic comparative methods (though currently limited to one continuous and one discrete feature) may allow a more satisfactory inclusion of continuous data (Boyko et al., 2023) but have yet to be applied to AMR.

Other directions for future research involve the properties of the source data (Diaz-Uriarte & Johnston, 2024; Schill et al., 2024). For resistance phenotypes, evolution between binary states of susceptible/resistance markers is reasonably intuitive. But for genetic features, there may be a large collection of variants that have similar consequences but are unlikely to ever all be accumulated in a given strain. Genomic AMR datasets often exhibit pronounced sparsity, with large numbers of relevant features but only a small number ever acquired by a given strain (Holt et al., 2015). Can such sparse datasets be transformed – for example, by clustering similar genes or other resistance features (Dauda et al., 2024) – so that inference using the transformed data is more informative?

Uncertainty in the source data – whether a feature is really present or absent in a sample – is also a pertinent feature in AMR studies. Observations of AMR features (phenotypic or genotypic) may be incomplete, due to (for example) only limited panels of drugs used in phenotyping, or incomplete sequencing. Uncertainty can also be involved in the classification of features which are measured, particularly as the functions of, and relationships between, "resistance genes" (and other upstream/downstream genes) remain incompletely understood. In addition to the uncertainty, the acquired resistance we are trying to model might affect the observability of the different events or combinations of events we are trying to model, leading to possible collider bias – an issue addressed in (Schill et al., 2024). Several EvAM approaches allow for incomplete or uncertain cross-sectional observations (Gerstung et al., 2009; Montazeri et al., 2016; Williams et al., 2013), but uncertainty in EvAM approaches with phylogenetically-embedded data – where ancestral states are also uncertain – is less well developed (Aga et al., 2024; Gao et al., 2022; Greenbury et al., 2020) and an important target for future research.

What if no phylogeny is available to connect the observations? One approach is to take two limits. First, incorrectly assume that observations are independent, and perform inference in this setting (subject to pseudoreplications). Second, estimate a phylogeny using a maximum parsimony assumption with the data itself as an alternative. Then, the maximum and minimum possible amounts of parallel evolution are captured by the two approaches, and the "true" picture will fall somewhere between the two. Then the outputs of the two alternatives pictures can be viewed as "bounds" on the likely underlying dynamics (Dauda et al., 2024).

## Conclusions

We have tried to summarise some of the ways that evolutionary accumulation modelling may contribute useful insights in the study of AMR. EvAM approaches are a different kind of tool to the approaches used to cluster, link, and analyse genomic and metagenomic sequence data – they consider how discrete, AMR-related features evolve across the phylogeny that connects a collection of isolates. They have substantial power to learn the evolutionary pathways by which these features emerge, and to further analyse interactions between features and predict future behaviours. We hope that some of these aspects of EvAM may be of interest and potential use for both basic and clinical AMR researchers, and hope to convince this community that EvAM approaches can form a valuable part in data-driven research on AMR evolution.

## Acknowledgements

This work was supported by the Trond Mohn Foundation [project HyperEvol under grant agreement No. TMS2021TMT09 to IGJ], through the Centre for Antimicrobial Resistance in Western Norway (CAMRIA) [TMS2020TMT11]. This project has received funding from the European Research Council (ERC) under the European Union's Horizon 2020 research and innovation programme [grant agreement No. 805046 (EvoConBiO) to IGJ]. This research was supported by grant PID2019-111256RB-I00 funded by MCIN/AEI/10.13039/501100011033 to Ramon Diaz-Uriarte.